\documentclass[cits]{PoS}

\usepackage{graphicx}
\usepackage{amsmath}

\newcommand{\eref}[1]{Eq.~(\ref{#1})}
\newcommand{\fref}[1]{Fig.~\ref{#1}}

\title{On propagators and vertices of Landau gauge Yang-Mills theory}

\ShortTitle{On propagators and vertices of Landau gauge Yang-Mills theory}

\author{\speaker{Markus Q. Huber}\\
        Theoriezentrum, Institut f\"ur Kernphysik, TU Darmstadt,
        64289 Darmstadt, Germany\\
        E-mail: \email{markus.huber@physik.tu-darmstadt.de}}

\author{Adrian Lorenz Blum\\
        Theoriezentrum, Institut f\"ur Kernphysik, TU Darmstadt,
        64289 Darmstadt, Germany\\
        E-mail: \email{adrian@theorie.ikp.physik.tu-darmstadt.de}}

\author{Mario Mitter\\
	Institut f\"{u}r Theoretische Physik, Ruprecht-Karls-Universit\"{a}t Heidelberg, 69120 Heidelberg, Germany\\
        E-mail: \email{m.mitter@thphy.uni-heidelberg.de}}

\author{Lorenz von Smekal\\
        Theoriezentrum, Institut f\"ur Kernphysik, TU Darmstadt,
        64289 Darmstadt, Germany\\
        Institut f\"ur Theoretische Physik, Justus-Liebig-Universit\"at,
        35392 Giessen, Germany\\
        E-mail: \email{lorenz.smekal@physik.tu-darmstadt.de}}

\abstract{
We calculate the three-point functions of pure Landau gauge QCD and investigate their influence on the propagators. As expected, the ghost-gluon vertex leads only to minor modifications, while the three-gluon vertex has a sizeable impact on the mid-momentum regime of the gluon propagator. We describe an effective model of the three-gluon vertex that includes contributions from the neglected two-loop diagrams and thus allows to obtain propagators in good agreement with lattice results. We also determine the three-gluon vertex from these propagators and find good agreement with lattice results as well. In turn, these results allow us to assess the effect of the missing two-loop diagrams in the gluon propagator equation. Finally, we present the first self-consistent calculation that includes all two-and three-point functions.
}

\FullConference{From quarks and gluons to hadronic matter: A bridge too far?\\
		 2-6 September, 2013\\
		 European Centre for Theoretical Studies in Nuclear Physics and Related Areas (ECT*), Villazzano, Trento (Italy)}

\begin{document}

\section{Introduction}

Functional methods are powerful tools to investigate quantum field theories beyond the perturbative regime. In quantum chromodynamics (QCD) they can be applied, for example, for the calculation of (pseudo-)order parameters, see, e.g., \cite{Fischer:2009wc,Braun:2009gm,Fischer:2012vc,Hopfer:2012qr,Fischer:2013eca}, or the description of hadronic properties, see, for instance, \cite{Alkofer:2000wg,Eichmann:2013afa} and refs. therein. A basis for this progress is a solid understanding of the behavior of correlation functions in the vacuum. Here we report on recent developments in the calculation of QCD Green functions from Dyson-Schwinger equations (DSEs) at zero temperature and without quarks.

The calculation of Green functions is technically demanding. Mostly, one resorts to Monte-Carlo simulations on discretized space-time lattices or uses functional methods. The former require fast computers, the latter insight on how to devise a good truncation. Enlarging truncations is a continuously ongoing process and leads to larger and more complicated systems of equations that are solved. A successful approach is also the combination of both, viz. to use lattice results as input for functional equations. This can reduce the truncation errors in the latter and allows at the same time to calculate quantities that are very costly or not feasible at all on the lattice.

When a truncation for a system of DSEs is devised, the asymptotic regimes provide good guidelines. However, in between, i.e., around the non-perturbative scale $\Lambda_{QCD}$ up to a few $GeV$, it is hard to quantitatively assess the magnitude as well as the source of the error. Typical sources of quantitative deviations are neglected diagrams and the way higher Green functions are treated. To give a well-known example, the gluon dressing function as calculated from the system of the two propagators, see, for example, \cite{vonSmekal:1997is,vonSmekal:1997vx,Pawlowski:2003hq,Fischer:2002hn,Aguilar:2008xm,Fischer:2008uz,Pennington:2011xs,LlanesEstrada:2012my}, has compared to lattice data \cite{Cucchieri:2007md,Sternbeck:2007ug,Bogolubsky:2009dc} a deficiency of approximately $20\,\%$ at $1\,GeV$ that is attributed to effects of the three-gluon vertex, which is modeled, and the neglected two-loop diagrams. The importance of higher vertices is confirmed by calculations with the functional renormalization group \cite{Fischer:2008uz}, where the gap to lattice results disappears after taking into account the dressed three- and four-gluon vertices. We will pinpoint the source of the gap between the DSE gluon propagator and corresponding lattice results in Sec.~\ref{sec:3g} by comparing propagators obtained with a lattice-compatible three-gluon vertex with the corresponding lattice results. On the other hand, we also solve the complete system of DSEs up to three-point functions. While all quantities that had to be modeled in most earlier calculations, namely the ghost-gluon vertex and the three-gluon vertex, are included dynamically now, the four-gluon vertex enters for the first time. One of our main findings is that for a quantitative, self-consistent description of Green functions from DSEs the two-loop diagrams in the gluon propagator DSE must be included, since our result for the three-gluon vertex cannot provide all of the missing strength in the mid-momentum regime. Note that first calculations of two-loop terms exist \cite{Bloch:2003yu,Mader:2013ru}, but no definite conclusion on the magnitude of their effect could be drawn. One complication in such calculations are the nested and overlapping divergences appearing for diagrams with more than one loop.

Finally, the three-gluon vertex is not only of interest as an ingredient for the gluon propagator DSE. In fact, it is a bottleneck for several applications. For example, in order to go beyond the rainbow-ladder truncation, which is kind of standard in bound state calculations with functional methods, the three-gluon vertex is required, e.g., \cite{Fischer:2009jm}, and also for the coupling of the gluonic to the matter sector via the quark-gluon vertex, see, e.g., \cite{Alkofer:2008tt,Hopfer:2013np}. Given that data for the three-gluon correlation is important for many phenomenological applications, the interest in this quantity was rekindled in recent years, see, for instance, \cite{Cucchieri:2008qm,Maas:2011se} for lattice results, \cite{Alkofer:2008dt,Aguilar:2013vaa} for analytic considerations, \cite{Pelaez:2013cpa} for a perturbative calculation with a mass term for the gluon field and \cite{Huber:2012kd} for a leading order DSE calculation.

Before we present our results for the three-gluon vertex we will explain the truncation scheme in Sec.~\ref{sec:trunc_scheme} and shortly describe in Sec.~\ref{sec:props+ghg} how we obtain the propagators used as input. They are taken from ref.~\cite{Huber:2012kd} where the model for the three-gluon vertex was adjusted such as to effectively include also two-loop contributions in the gluon propagator DSE. Note that this was just the means to an end, namely to obtain a quantitatively good description of the two-point correlations. This model vertex, although qualitatively respecting many properties of the vertex, does not represent the real vertex, which is calculated in Sec.~\ref{sec:3g}. On the other hand, given the reasonable agreement we obtain for the calculated vertex with lattice data, we can test it in the gluon propagator DSE, see Sec.~\ref{sec:gl_dressing}. Finally, we also solve the fully coupled system in Sec.~\ref{sec:selfConsistent}. For all calculations the \textit{Mathematica} package \textit{DoFun} \cite{Huber:2011qr,Alkofer:2008nt} was used for the derivation of the equations, which were then solved using the framework of \textit{CrasyDSE} \cite{Huber:2011xc}.

\section{Truncation scheme}
\label{sec:trunc_scheme}

\begin{figure}[tb]
  \includegraphics[width=\textwidth]{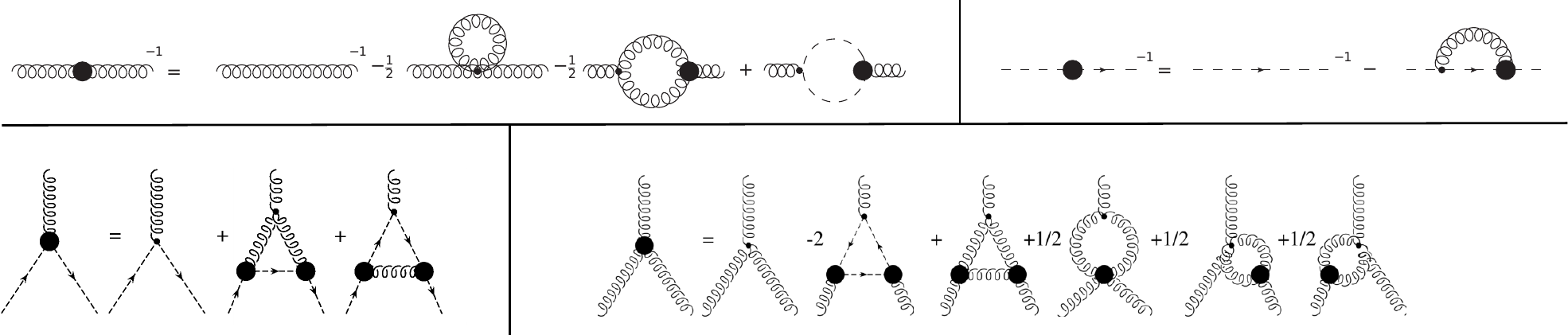}
  \caption{\label{fig:system_DSEs}The system of truncated DSEs. All internal propagators are dressed. Thick blobs denote dressed vertices. Wiggly (dashed) lines are gluons (ghosts).}
\end{figure}

Our truncation prescription for a given DSE is to keep all diagrams contributing perturbatively at one-loop order, i.e., we discard all two-loop diagrams together with the one-loop diagrams containing non-primitively divergent vertices like the ghost-gluon four-point function. Apart from the UV leading diagrams, our truncation contains all IR leading diagrams and furthermore the ghost propagator DSE stays unaffected. As usual three-point functions are required for the evaluation of the propagators and we will use models as well as self-consistent solutions in the following. In the DSE of the three-gluon vertex we encounter the four-gluon vertex which is the only quantity that is not determined self-consistently at the moment. With our truncation prescription, however, the inclusion of the four-gluon vertex would lead to a closed system and no new input would be required. The system of truncated DSEs is depicted in \fref{fig:system_DSEs}.

\section{Two-point functions and ghost-gluon vertex}
\label{sec:props+ghg}

First, we discuss the system of propagators and ghost-gluon vertex. Note that there are two DSEs for the ghost-gluon vertex and we choose the one where the gluon leg is connected to the bare vertices rather than a ghost leg as this reduces the dependence on the three-gluon vertex model. The dressing function for the ghost propagator is denoted by $G(p^2)$, that for the gluon propagator by $Z(p^2)$. In Landau gauge, it is sufficient to use one dressing function for the ghost-gluon vertex, $A(k^2, p^2, \alpha)$, where $k$ is the gluon and $p$ the anti-ghost momentum and $\alpha$ the angle between the two. The corresponding DSEs were solved self-consistently for the first time in two dimensions \cite{Huber:2012zj} and later in four dimensions \cite{Huber:2012kd}. 

\begin{figure}[tb]
 \begin{center}
  \includegraphics[width=0.49\textwidth]{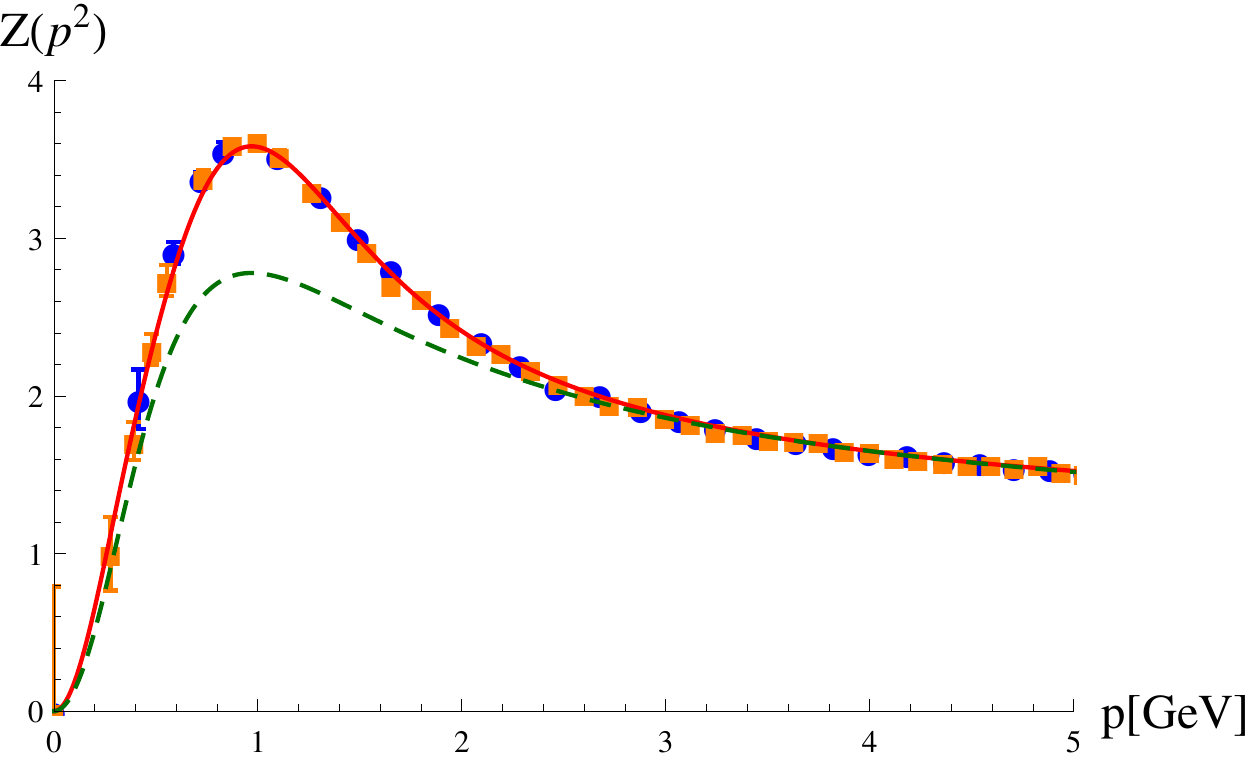}
  \includegraphics[width=0.49\textwidth]{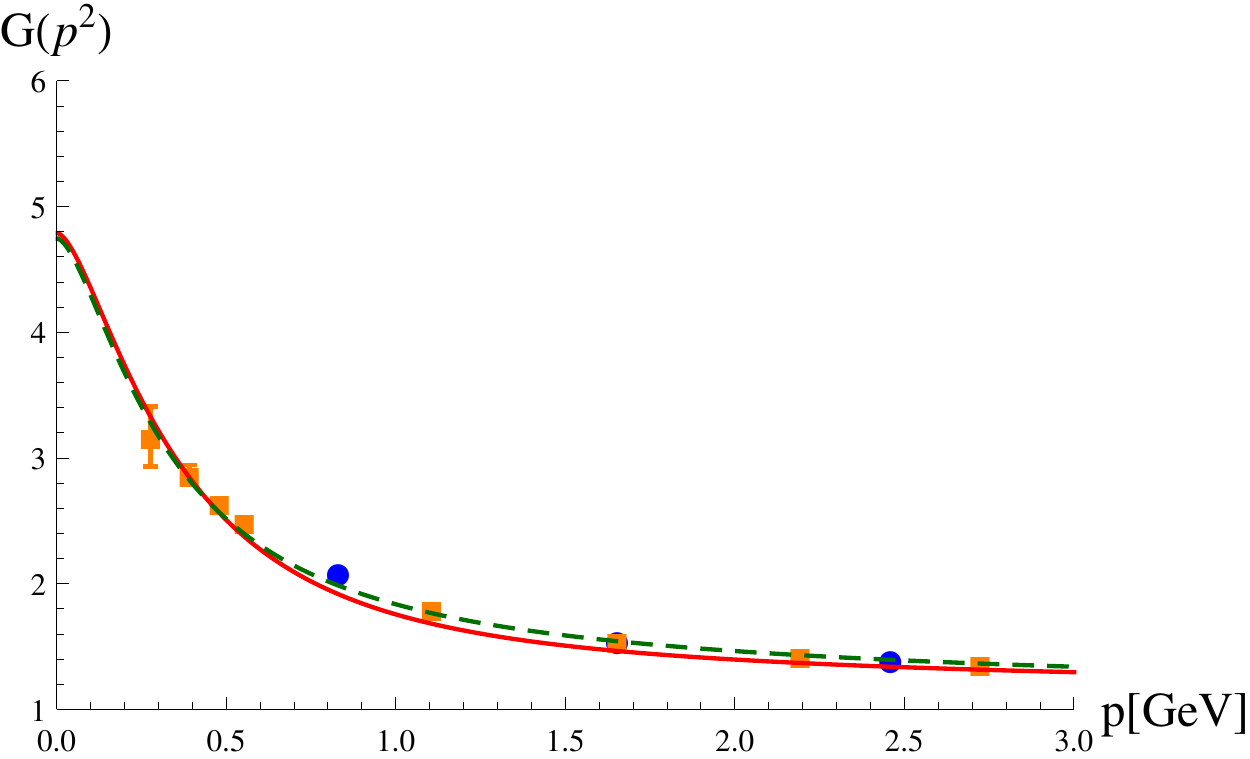}
  \caption{\label{fig:props_optEff3g}The $SU(3)$ gluon and ghost dressing functions $Z(p^2)$ and $G(p^2)$ in comparison with lattice data \cite{Sternbeck:2006rd}. The red/continuous lines represent the results with a dynamic ghost-gluon vertex and the optimized effective three-gluon vertex, the green/dashed lines a reference calculation with a bare ghost-gluon vertex and the three-gluon vertex of ref.~\cite{Fischer:2002eq}.}
 \end{center}
\end{figure}

\begin{figure}[tb]
 \begin{center}
  \includegraphics[width=0.49\textwidth]{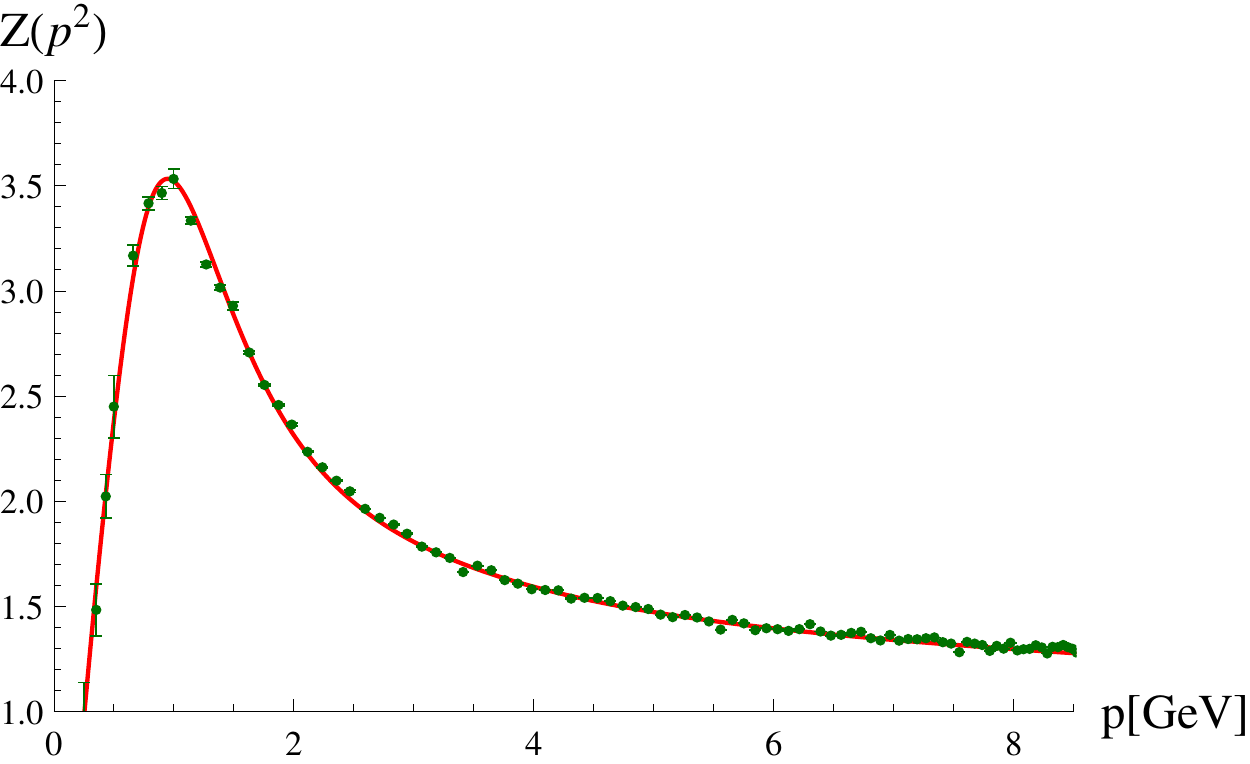}
  \includegraphics[width=0.49\textwidth]{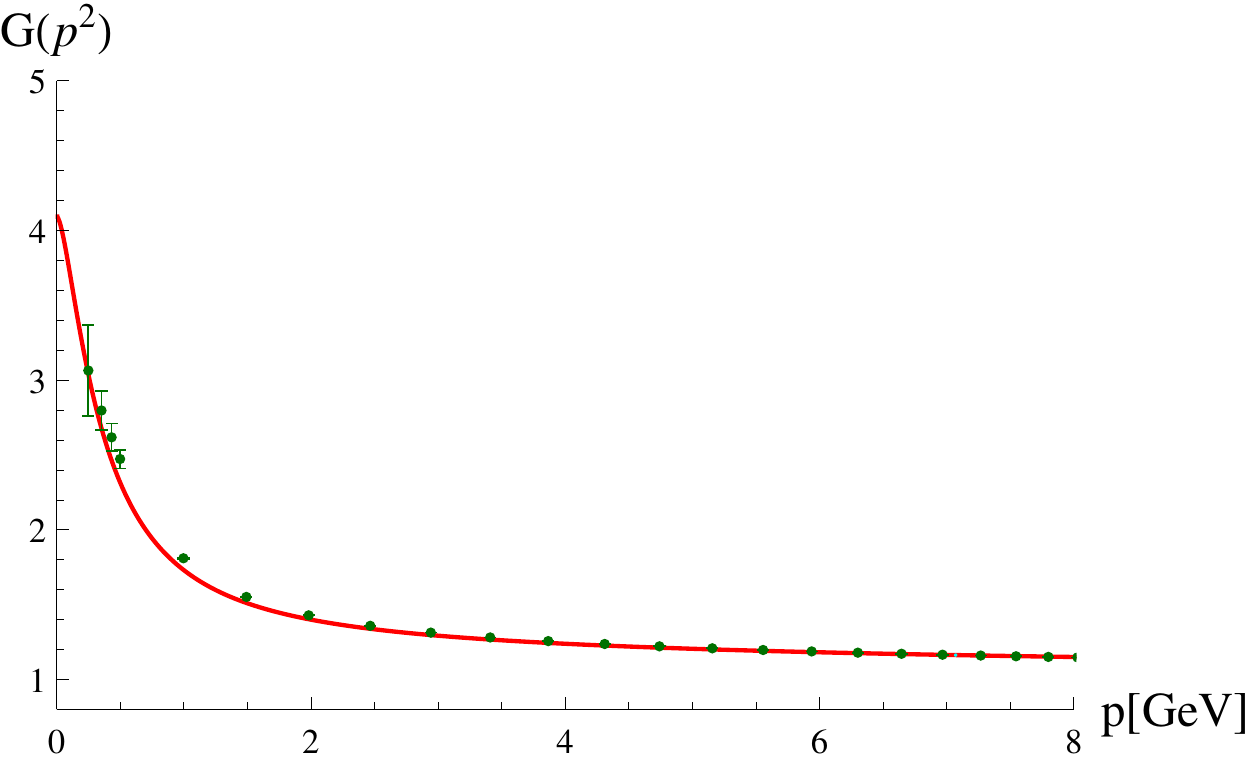}
  \caption{\label{fig:props_optEff3g-SU2}The gluon and ghost dressing functions $Z(p^2)$ and $G(p^2)$ for $SU(2)$ in comparison with lattice data \cite{Sternbeck:2007ug}. The red/continuous lines represent the results with a bare ghost-gluon vertex and the optimized effective three-gluon vertex.}
 \end{center}
\end{figure}

Within this system the only unknown quantity is the three-gluon vertex which is modeled. Taking as prerequisites that this model reflects the Bose symmetry of the vertex and has its anomalous dimension we start with
\begin{align}
\label{eq:3g_model_UV}
 D^{A^3,UV}(x,y,z)&=G\left(\frac{x+y+z}{2}\right)^{\alpha_{3g}}Z\left(\frac{x+y+z}{2}\right)^{\beta_{3g}}
\end{align}
where $\alpha_{3g}=-17/9$ and $\beta_{3g}=0$ for the decoupling type solution and $x$, $y$ and $z$ are the squares of the external momenta. As it turned out, \eref{eq:3g_model_UV} describes the UV behavior of the vertex quite well so that we use this expression for the extrapolation in the UV.
In two and three dimensions it was found in lattice calculations that the projection for the three-gluon vertex as employed in those calculations, see \eref{eq:tg_proj} below, features a zero crossing. A full calculation of the three-gluon vertex DSE in two dimensions \cite{Huber:2012zj} and a leading order calculation in three dimensions \cite{Campagnari:2010wc} confirmed this. Also in four dimensions such a zero crossing exists which is confirmed by a leading order DSE calculation \cite{Huber:2012kd}, by analytic considerations for the specific kinematic configuration where one gluon momentum vanishes \cite{Aguilar:2013vaa} as well as by a perturbative calculation with a massive gluon propagator \cite{Pelaez:2013cpa}. In \cite{Blum:2014gna} it was shown that this feature is retained in a fully self-consistent solution. In the used model the zero crossing can be implemented by adding the term
\begin{align}
  D^{A^3,IR}(x,y,z)&=h_{IR} \,G(x+y+z)^{3}(f^{3g}(x)f^{3g}(y)f^{3g}(z))^4 \quad \text{with}\quad   f^{3g}(x):=\Lambda^2_{3g}/(\Lambda_{3g}^2+x).
\end{align}
The full vertex model is then given by
\begin{align}
 D^{A^3}(x,z,y)=D^{A^3,IR}(x,y,z)+D^{A^3,UV}(x,y,z).
\end{align}
The parameters of this model can be utilized to modify the strength of the gluon-loop in the gluon propagator DSE. Typically $h_{IR}$ is chosen as $1$ and the scale parameter $\Lambda_{3g}$ is varied. It can be chosen such that the gap between lattice and DSE propagators is closed. We call this vertex the optimized effective three-gluon vertex as it effectively includes two-loop contributions in the gluon propagator DSE \cite{Huber:2012kd}. The good agreement between lattice and DSE results is shown in \fref{fig:props_optEff3g}. For more details we refer to ref.~\cite{Huber:2012kd} or ref.~\cite{Huber:2013xb}. The Schwinger function of the gluon, which clearly indicates violation of positivity, can also be found in the latter reference. This model of the three-gluon vertex can be used to achieve good agreement with lattice results even in a truncation with a bare ghost-gluon vertex which is explicitly demonstrated for $SU(2)$ in \fref{fig:props_optEff3g-SU2}.

\begin{figure}[tb]
\includegraphics[width=0.5\textwidth]{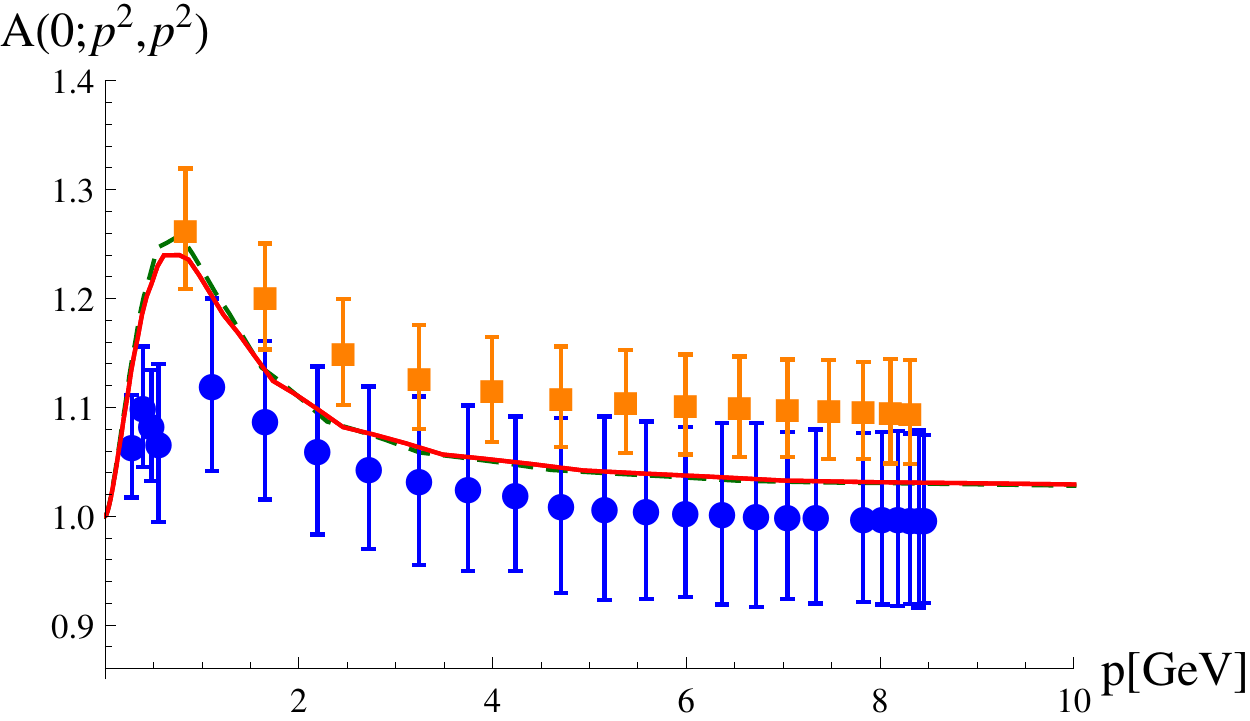}
\includegraphics[width=0.5\textwidth]{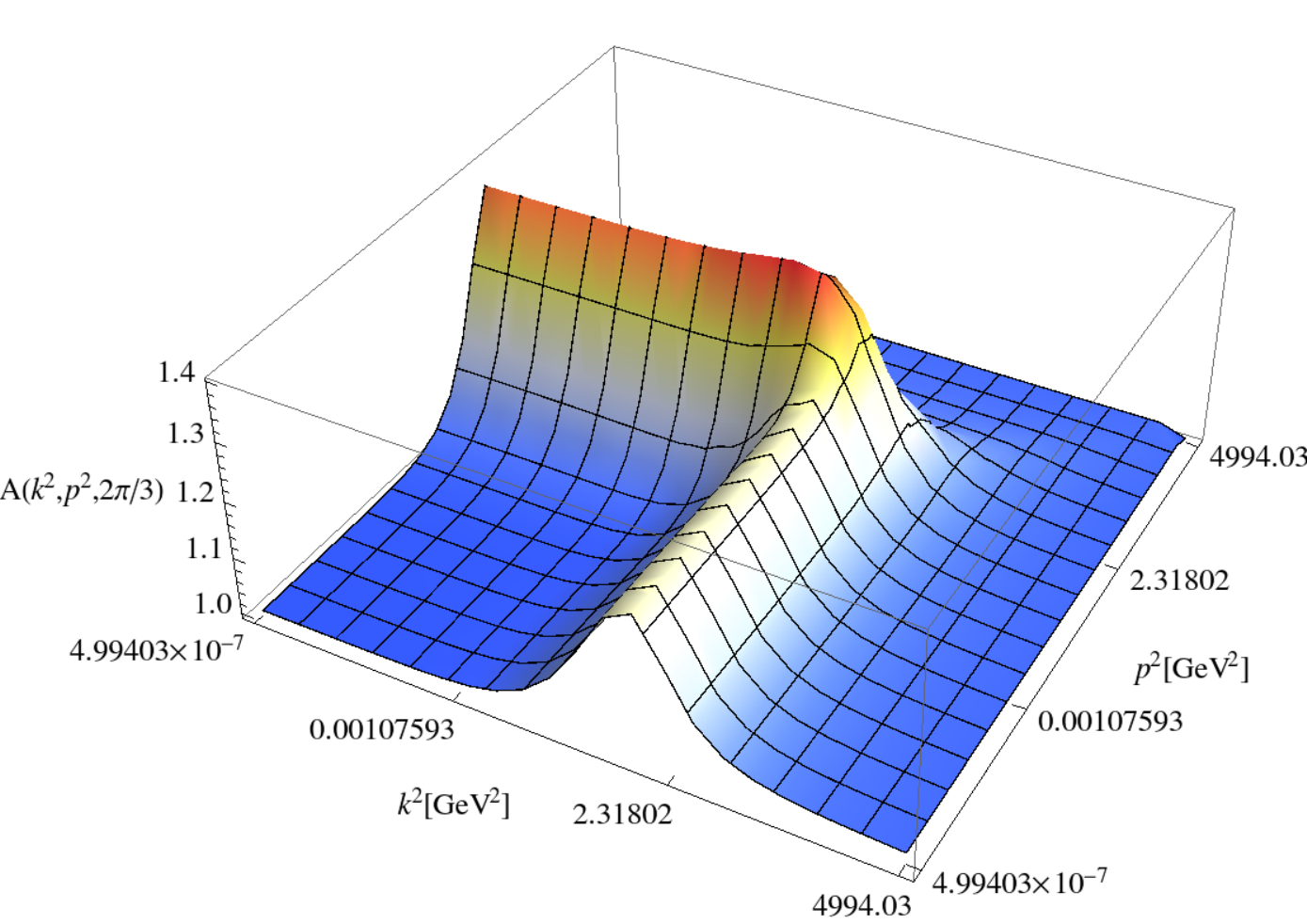}
\caption{\label{fig:ghg}\textit{Left}: The $SU(3)$ ghost-gluon vertex dressing function at vanishing gluon momentum. The two lines correspond to results from different three-gluon vertex models. The Lattice data is taken from \cite{Sternbeck:2006rd}. \textit{Right}: Ghost-gluon vertex dressing with angle $2\pi/3$ between the gluon and the anti-ghost momentum.}
\end{figure}

The ghost-gluon vertex was obtained with its full kinematic dependence. In \fref{fig:ghg} we compare the kinematic configuration of vanishing gluon momentum to corresponding lattice data. Within the error bars the results agree well with the lattice results. In the three-dimensional plot one sees that the bump in the mid-momentum region is higher if the gluon momentum is smaller than the anti-ghost momentum. This is also reflected in lattice data for $SU(2)$ \cite{Cucchieri:2008qm}.

Actually, the bump in the ghost-gluon vertex can already be observed in a semi-perturbative calculation where bare vertices and dressed propagators are used \cite{Schleifenbaum:2004id}. Later this setup was improved by solving the vertex DSE together with the ghost propagator DSE \cite{Aguilar:2013xqa} where the momentum dependence of the vertex was approximated with the gluon momentum scale and only bare vertices were used on the right-hand side of the vertex DSE. Also with a perturbative calculation using a massive gluon propagator the qualitative features of the lattice results can be reproduced \cite{Pelaez:2013cpa}. With the functional renormalization group the vertex was calculated at the symmetric point in ref.~\cite{Fister:2011uw}. Finally, the vertex can be described by an OPE motivated model \cite{Boucaud:2011eh} whose parameters were determined from lattice results in ref.~\cite{Dudal:2012zx}.

\section{The three-gluon vertex}
\label{sec:3g}

The lattice data available for the three-gluon vertex \cite{Cucchieri:2008qm} is unfortunately not suitable to restrain models for the vertex much. For example, the zero crossing has not even been seen yet in four dimensions on the lattice and according to analytic estimates and explicit calculations quite large lattices will be required for this \cite{Aguilar:2013vaa, Blum:2014gna}. Here, we calculate the vertex from its DSE using as input the propagators described in the previous section. Given the agreement of the propagators with lattice data, any errors that are found must be due to the neglected diagrams or tensors.

In comparison to the ghost-gluon vertex we have the complication that the vertex is not UV finite. For its renormalization we have to remember that its renormalization constant $Z_1$ is fixed by a Slavnov-Taylor identity: $Z_1=\tilde{Z}_1 Z_3/\tilde{Z}_3$. We note that $\tilde{Z}_1=1$ in the used \textit{MiniMOM} renormalization scheme \cite{vonSmekal:2009ae} and the renormalization constants $Z_3$ and $\tilde{Z}_3$ can be calculated from the gluon and ghost propagators. $Z_1$ appears in the tree-level diagram of the three-gluon vertex DSE. Moreover, it also appears in front of loop diagrams, as does $Z_4$, the renormalization constant of the four-gluon vertex.
To prevent deviations from the one-loop resummed anomalous dimension, we use $Z_1=Z_3/\tilde{Z}_3$ only in the tree-level diagram and replace the renormalization constants in front of the loop diagrams by a momentum-dependent function \cite{vonSmekal:1997vx}. With this replacement we effectively add a renormalization group (RG) improvement that allows us to recover the correct anomalous dimension of one-loop resummed perturbation theory. The corresponding functions are given by 
\begin{align}
\label{eq:3g-RG}
 D^{A^3,RG}(p_i^2)&=G\left(\overline{p}^2\right)^{\alpha_{3g}}Z\left(\overline{p}^2\right)^{\beta_{3g}},\\
\label{eq:4g-RG}
 D^{A^4,RG}(p_i^2)&=G\left(\overline{p}^2\right)^{\alpha_{4g}}Z\left(\overline{p}^2\right)^{\beta_{4g}}
\end{align}
where $\overline{p}^2=\sum_{i}p_i^2/2$ and the $p_i$ are the external momenta. For the three-gluon vertex this is actually the same expression as the UV part of the model in \eref{eq:3g_model_UV}. Determining the exponents for the four-gluon vertex in the same way, we obtain $\alpha_{4g}=-8/9$ and $\beta_{4g}=0$ for the decoupling solution. Also in the gluon propagator equation such an RG improvement term is used \cite{Huber:2012kd}.

The three-gluon vertex has four transverse tensors. Here, however, we restrict ourselves to the tree-level part. For a discussion of the full basis see \cite{Alkofer:2013ip}. The DSE is projected in the same way as on the lattice, i.e., 
 \begin{align}\label{eq:tg_proj}
 D^{A^3}_{proj}(p^2,q^2, \alpha):=\frac{\Gamma_{\mu\nu\rho}^{A^3,abc,(0)}(p,q,r) P_{\mu\mu'}(p)P_{\nu\nu'}(q)P_{\rho\rho'}(r) \Gamma_{\mu'\nu'\rho'}^{A^3,abc}(p,q,r)}{\Gamma_{\alpha\beta\gamma}^{A^3,def,(0)}(p,q,r) P_{\alpha\alpha'}(p)P_{\beta\beta'}(q)P_{\gamma\gamma'}(r) \Gamma_{\alpha'\beta'\gamma'}^{A^3,def,(0)}(p,q,r)},
\end{align}
where $P_{\mu\nu}(p)$ is the transverse projector and $(0)$ indicates tree-level. The full vertex on the right-hand side of the DSE is then replaced as $\Gamma_{\mu\nu\rho}^{A^3,abc}(p,q,r) \rightarrow D^{A^3}_{proj}(p^2,q^2, \alpha)\Gamma_{\mu\nu\rho}^{A^3,abc,(0)}(p,q,r)$. Using this projection has the advantage that we can compare directly with lattice results and is useful when plugging the vertex results into the gluon propagator DSE as discussed below.

\begin{figure}[tb]
  \includegraphics[width=0.32\textwidth]{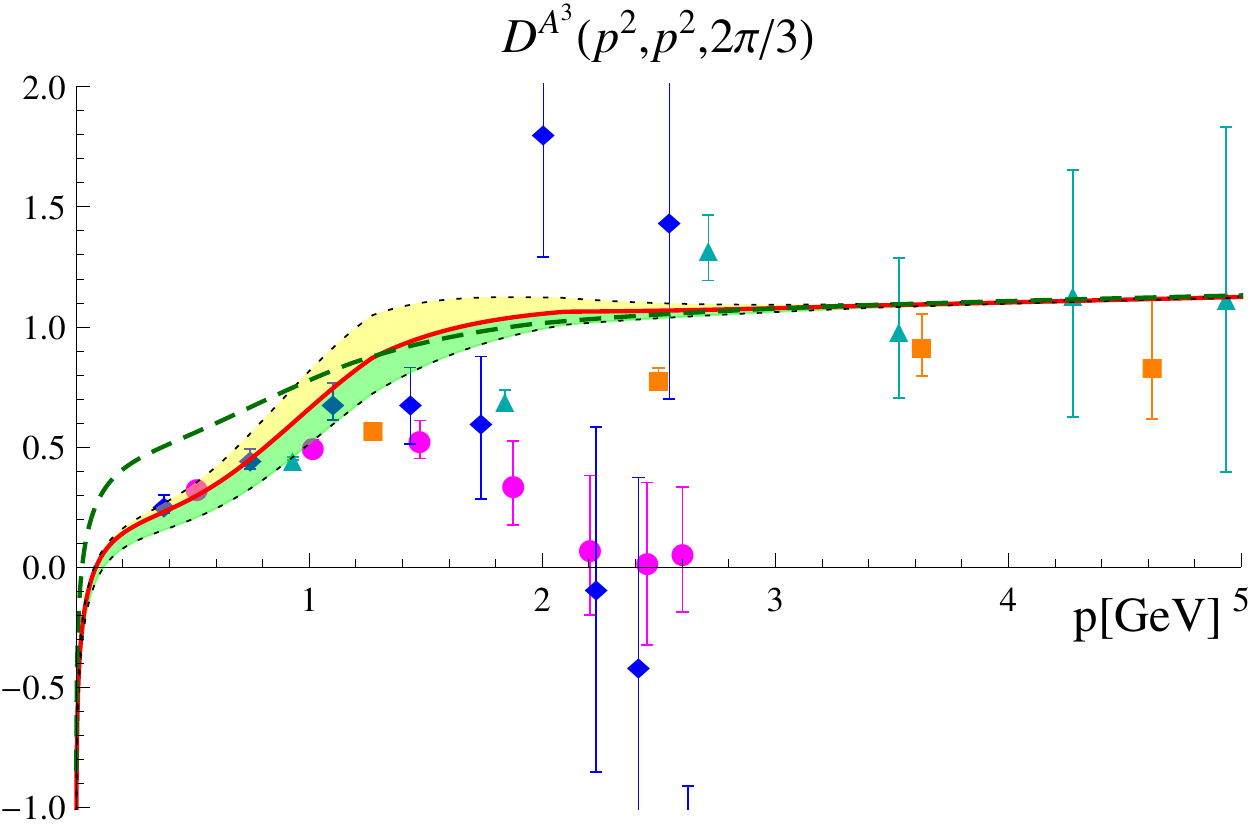}
  \includegraphics[width=0.32\textwidth]{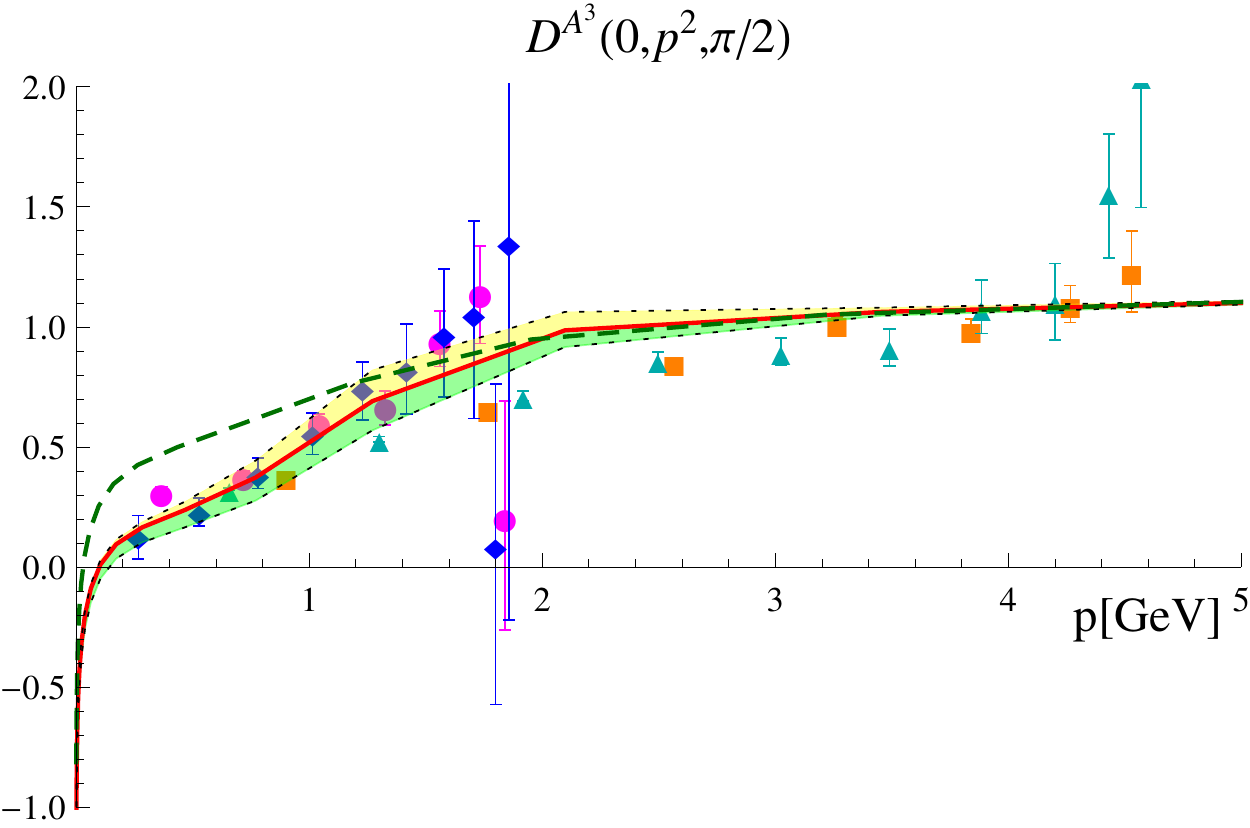}
  \includegraphics[width=0.32\textwidth]{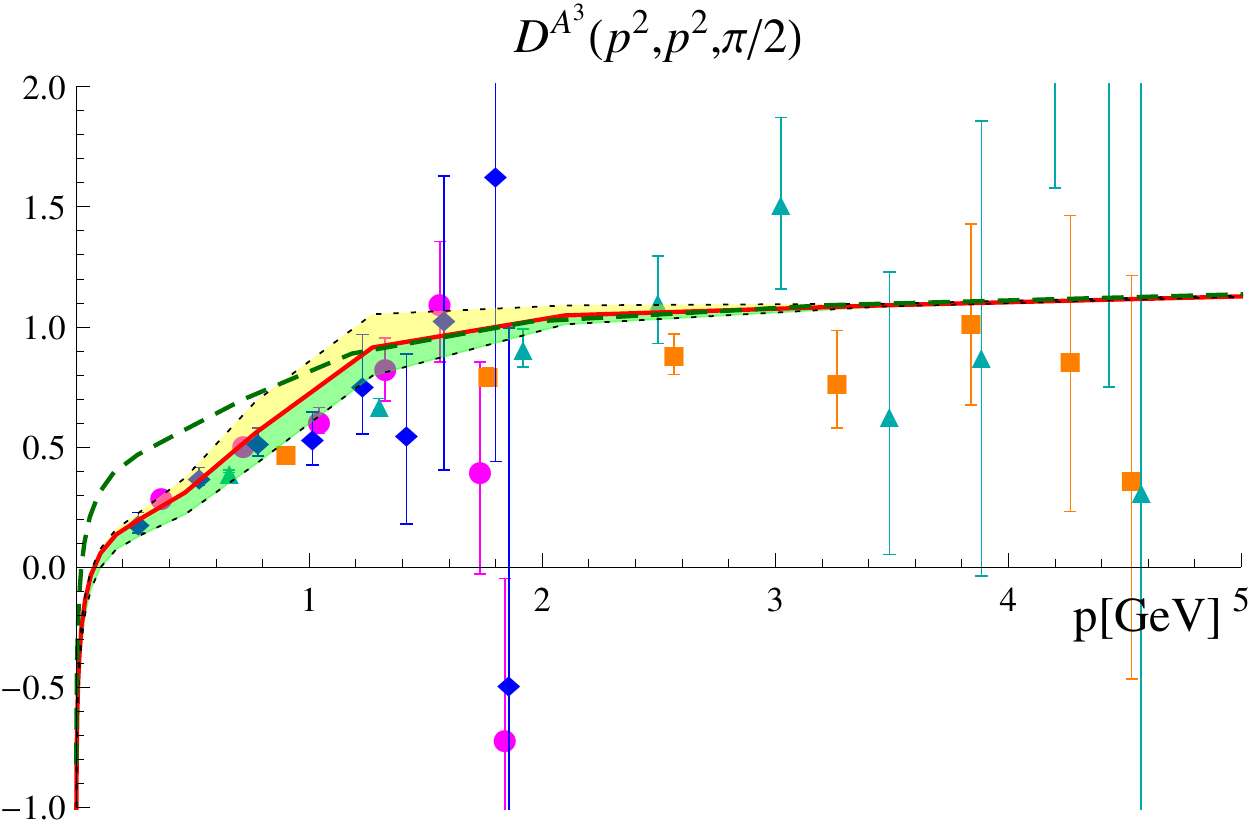}
  \caption{\label{fig:3g_singleScale}The dressing of the $SU(3)$ three-gluon vertex for three kinematical configurations (as denoted at the top of the plots) in comparison to lattice results for $N_c=2$ \cite{Cucchieri:2008qm}. Red/continuous line and band: $D^{A^3}_o(p^2,p^2,\alpha)$ with the following $a/b$: $1.5/1.95\,GeV^2$ (center), $1.5/1.46\,GeV^2$ (top), $2/1.95\,GeV^2$ (bottom). Green dashed line: From fully coupled system with $a=1.5$, $b=1.94\,GeV^2$.}
\end{figure}

\begin{figure}[tb]
  \includegraphics[width=0.32\textwidth]{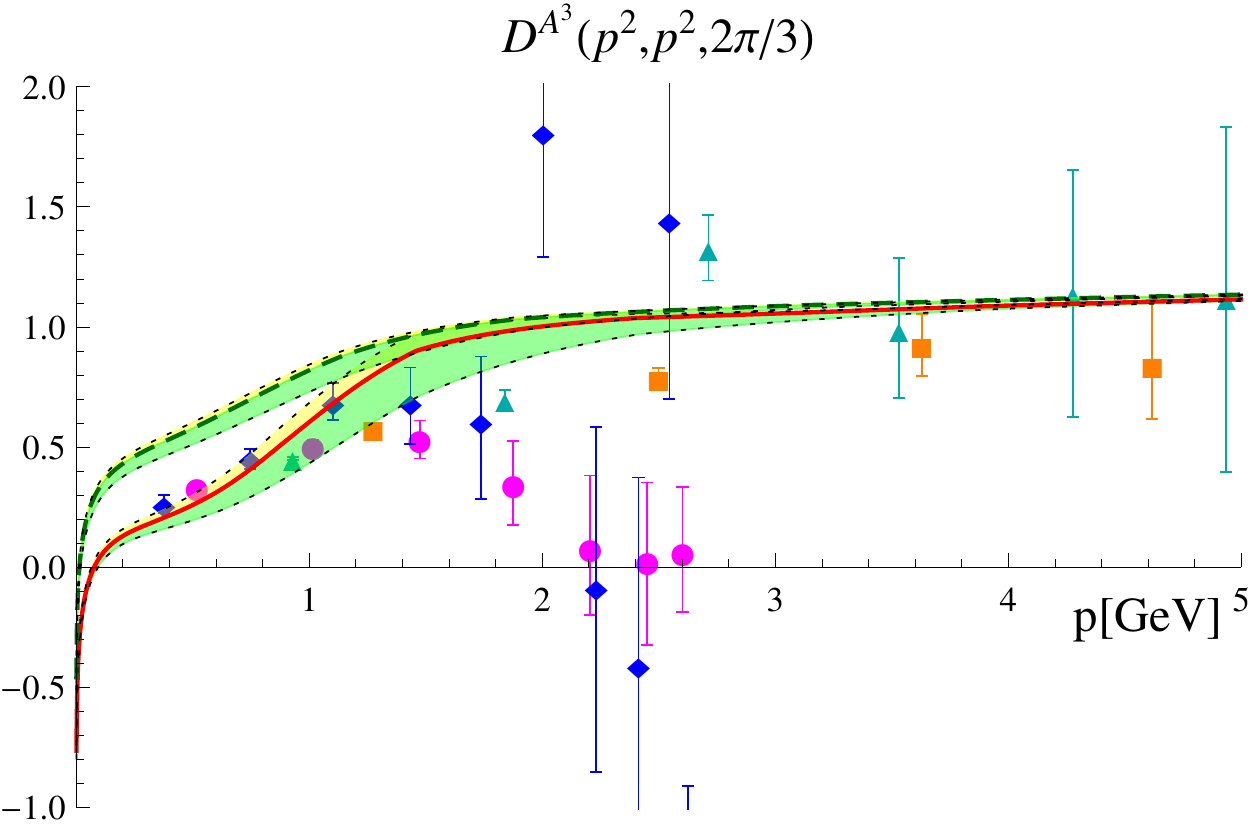}
  \includegraphics[width=0.32\textwidth]{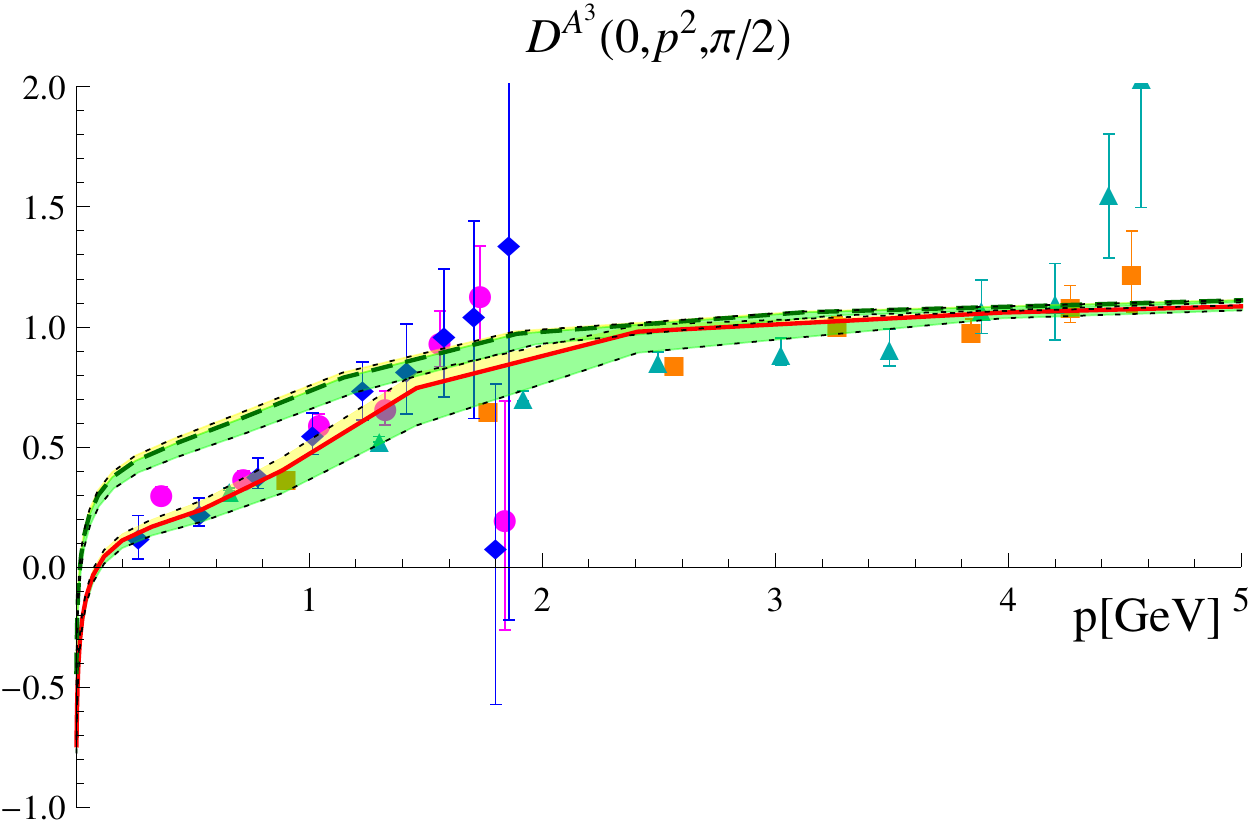}
  \includegraphics[width=0.32\textwidth]{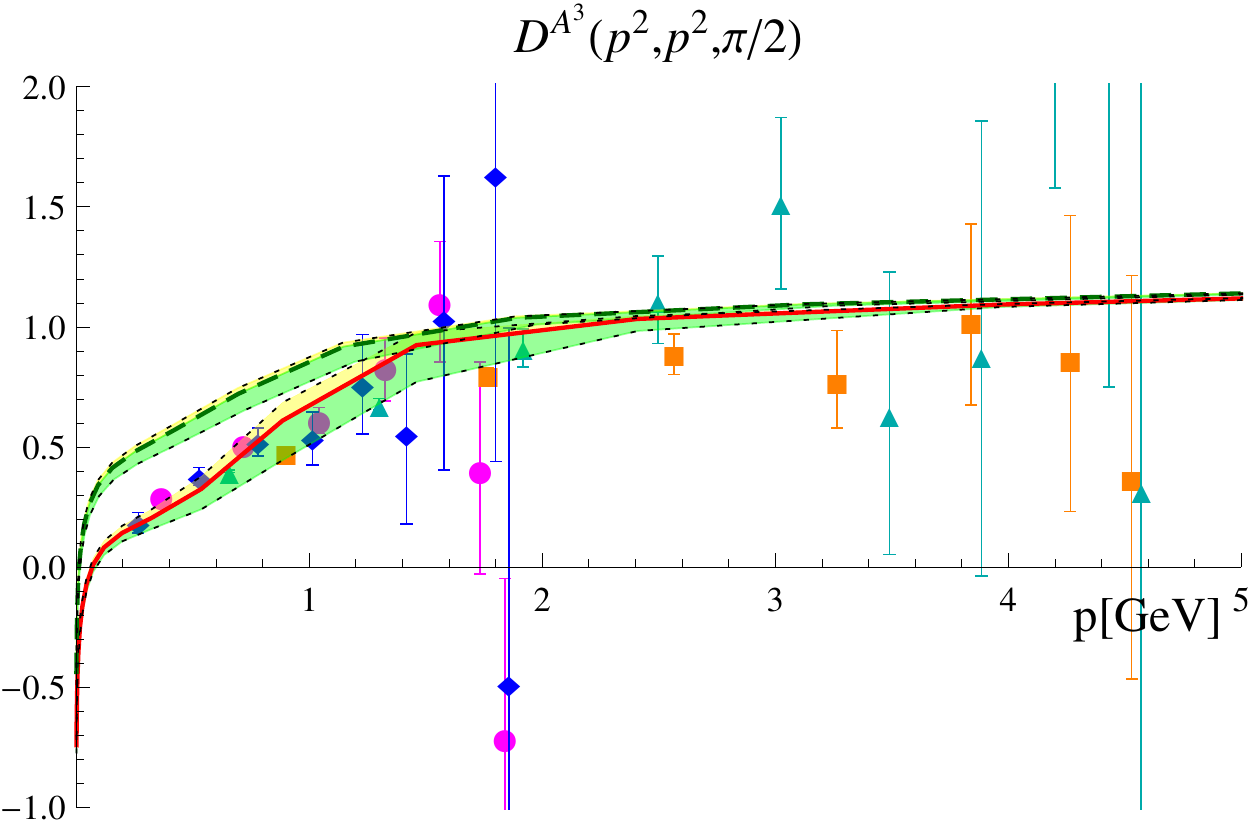}
  \caption{\label{fig:3g_singleScale-SU2}The dressing of the $SU(2)$ three-gluon vertex for three kinematical configurations (as denoted at the top of the plots) in comparison to lattice results for $N_c=2$ \cite{Cucchieri:2008qm}. Red/continuous line and (lower) band: $D^{A^3}_o(p^2,p^2,\alpha)$ with the following $a/b$ in the four-gluon vertex model: $1.5/1.67\,GeV^2$ (center), $1.25/1.67\,GeV^2$ (top), $1.5/3.34\,GeV^2$ (bottom). Green/dashed line and (upper) band: Fully coupled system. Same values for $a/b$.}
\end{figure}

\begin{figure}[tb]
  \includegraphics[width=0.45\textwidth]{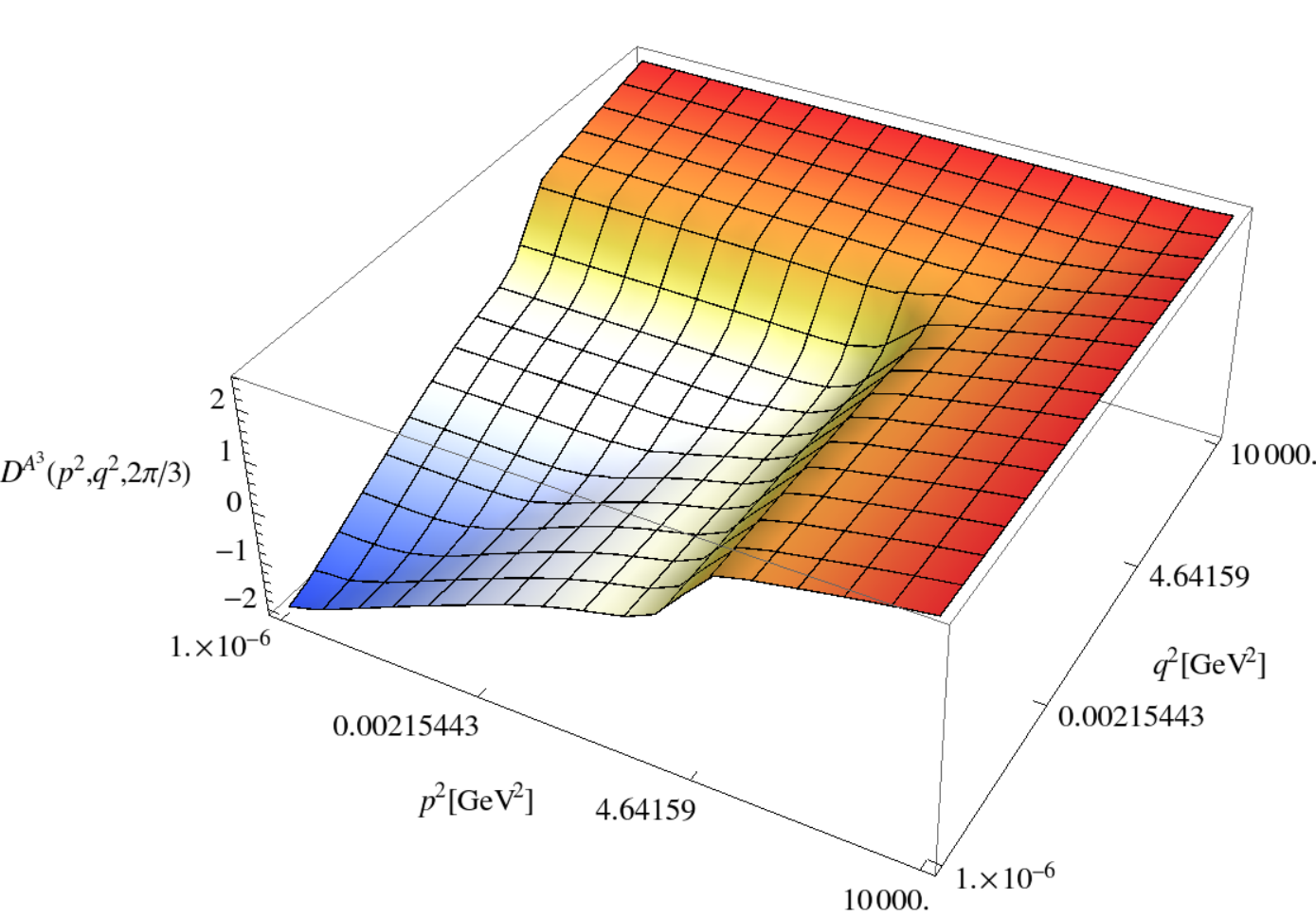}
  \hfill
  \includegraphics[width=0.45\textwidth]{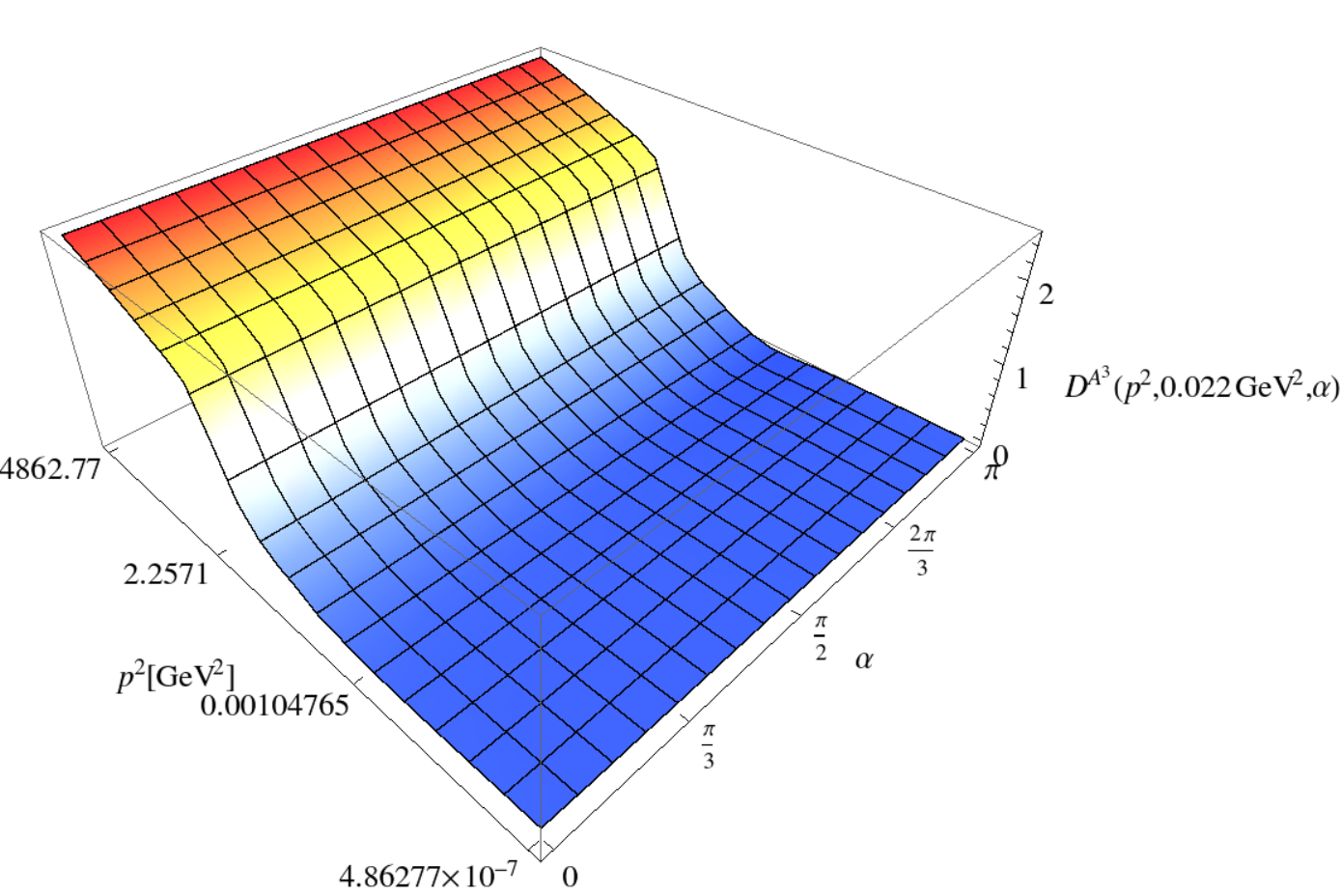}
  \caption{\label{fig:3g_3d}The $SU(3)$ three-gluon vertex dressing for fixed angle, $D^{A^3}_o(p^2,q^2,2\pi/3)$ (left), and for one fixed momentum, $D^{A^3}_o(p^2,0.022\,GeV^2, \alpha)$ (right).}
\end{figure}

We will first solve the three-gluon vertex, denoted in this setup by $D^{A^3}_o(p^2,q^2,\alpha)$, with the propagators from Sec.~\ref{sec:props+ghg}. This reduces the error from the input with the main uncertainties stemming from the following sources: Neglected diagrams, the reduced tensor basis and the model for the four-gluon vertex. As far as the tensor basis is concerned our results will show that the restriction to the tree-level is already a good approximation. On the other hand, the other error sources play a crucial role. The reason is that convergence of the calculation depends on the four-gluon vertex model. Considering the diagrams separately, one sees that the contribution from the gluon triangle features a bump in the mid-momentum regime \cite{Alkofer:2013ip}. This bump enhances itself in the iteration process if not counteracted by other contributions.\footnote{We thank G.~Eichmann and R.~Williams for bringing this to our attention, for numerical cross-checks, and for clarifying discussions on how to resolve the issue.} Whereas we cannot say if neglected diagrams are relevant for this, we found that the four-gluon vertex dressing is important. The RG-improvement, \eref{eq:4g-RG}, does not produce enough strength to counteract the bump in the gluon triangle but from a DSE analysis of the four-gluon vertex \cite{Kellermann:2008iw} we expect this model is inadequate in the mid-momentum regime anyway.
Thus we will adapt our model such that it contains additional strength in the regime up to a few $GeV$. Our ansatz for the four-gluon vertex is
\begin{align}\label{eq:4g-model}
  \Gamma^{A^4}(p, q, r, s)_{\mu\nu\rho\sigma}^{abcd} =(a \tanh(b/\bar{p}^2)+1)D^{A^4,RG}(p, q, r, s) \Gamma^{A^4}(p, q, r, s)_{\mu\nu\rho\sigma}^{abcd,(0)}.
\end{align}
The parameters $a$ and $b$ offer the possibility to vary the result of the three-gluon vertex DSE. Given the rather large error bars of the lattice results, we decided to use three sets of parameters. Thus we can produce a band of solutions that cover the lattice results, see figs.~\ref{fig:3g_singleScale} and \ref{fig:3g_singleScale-SU2} for $SU(3)$ and $SU(2)$, respectively. This will also allow  us to assess the effect of the four-gluon vertex on the gluon propagator DSE. Although the lattice data is for $SU(2)$, we use it for comparison with the $SU(3)$ results, since it is known from the propagators that the difference between $SU(2)$ and $SU(3)$ is rather small \cite{Sternbeck:2007ug}. The three-gluon vertex, shown in \fref{fig:3g_3d}, shows hardly any angle dependence. Finally we mention that the vertex is logarithmically IR divergent. This was found recently also within a perturbative calculation using a massive gluon propagator \cite{Pelaez:2013cpa} as well as by analytic calculations \cite{Aguilar:2013vaa} and agrees with earlier considerations about the decoupling solution \cite{Alkofer:2008jy}.

\begin{figure}[tb]
 \includegraphics[width=0.4\textwidth]{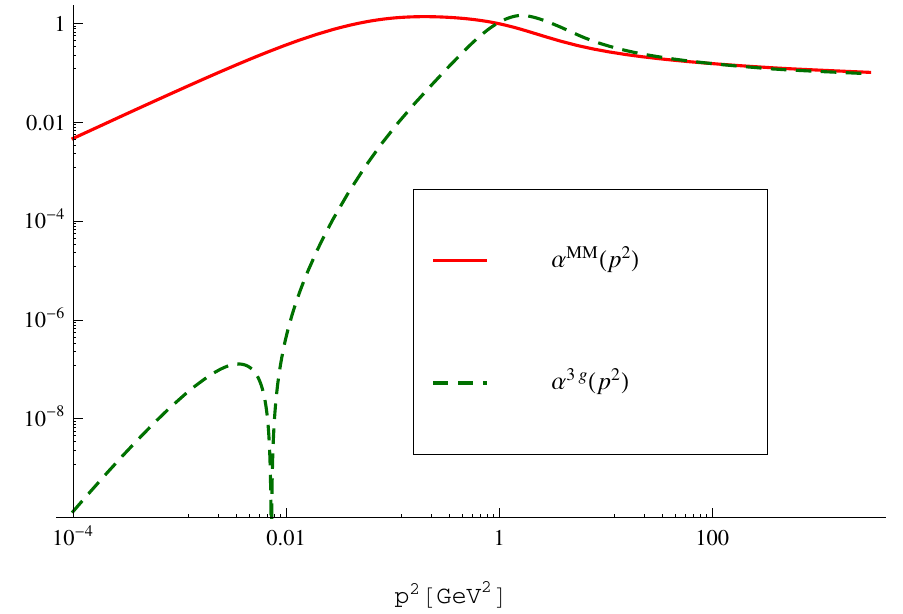}
 \hfill
 \includegraphics[width=0.4\textwidth]{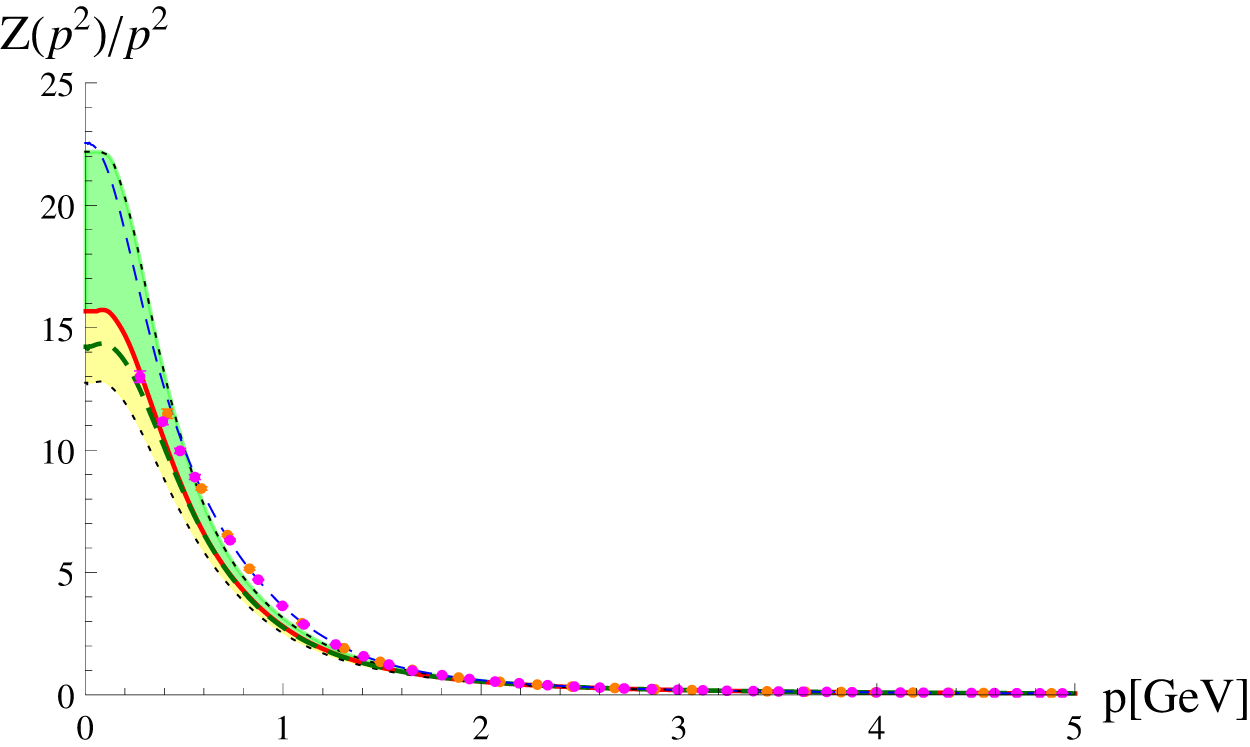}
  \caption{\label{fig:couplings}\textit{Left}: The couplings obtained from the $SU(3)$ ghost-gluon (red/continuous) and the three-gluon (green/dashed) vertices. \textit{Right}: The gluon propagators corresponding to the dressings in fig. 9.}%fig. number manual due to problem with style and using \ref in a caption
\end{figure}

Similar to the \textit{MiniMOM} coupling from the ghost-gluon vertex \cite{vonSmekal:1997is,vonSmekal:2009ae} one can calculate a coupling from the three-gluon vertex (and even higher Green functions) \cite{Alkofer:2004it}. It is given by
\begin{align}
  \alpha^{3g}(p^2)=\alpha(\mu^2)\frac{Z(p^2)^3 D^{A^3}\left(p^2,p^2, 2\pi/3\right)^2}{Z(\mu^2)^3 D^{A^3}\left(\mu^2,\mu^2, 2\pi/3\right)^2}.
\end{align}
The corresponding result from the three-gluon vertex in the middle of the band is shown in \fref{fig:couplings}. Due to the zero crossing the coupling touches zero. For zero momentum it vanishes.

\section{The gluon dressing function}
\label{sec:gl_dressing}

\begin{figure}[tb]
  \includegraphics[width=0.45\textwidth]{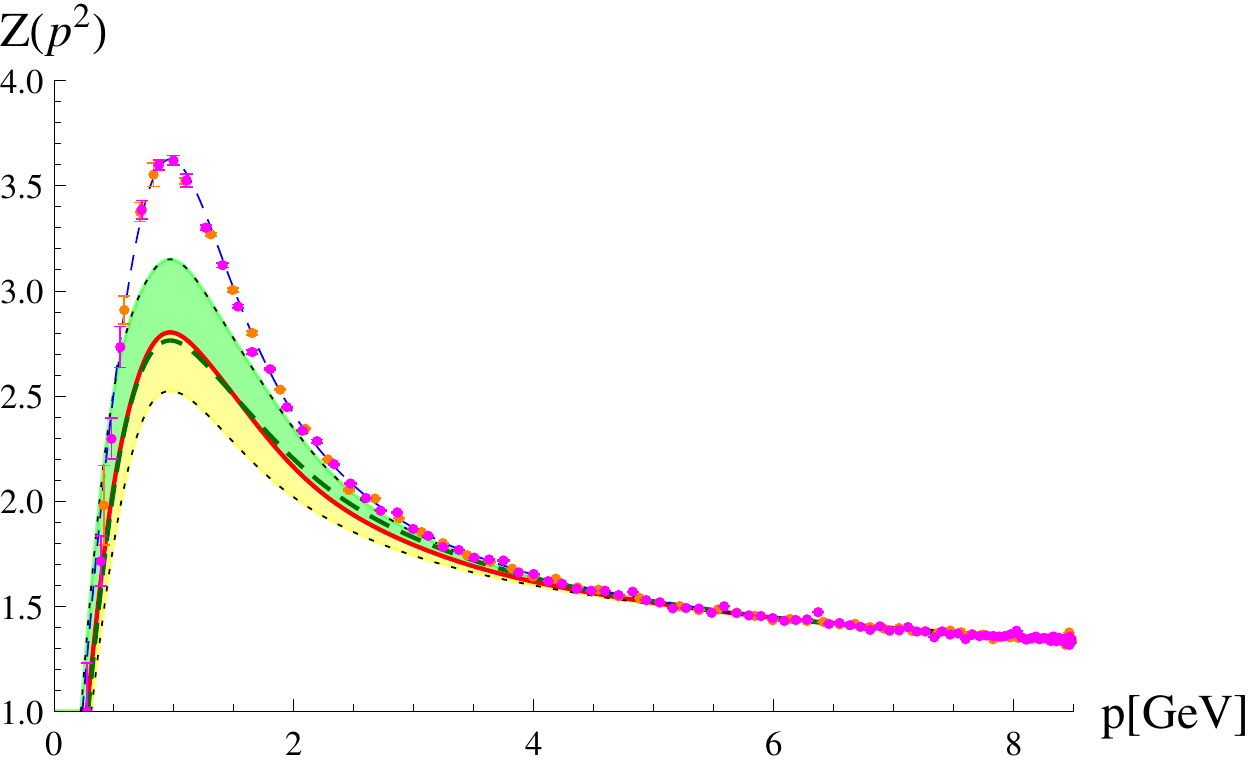}
  \hfill
\includegraphics[width=0.45\textwidth]{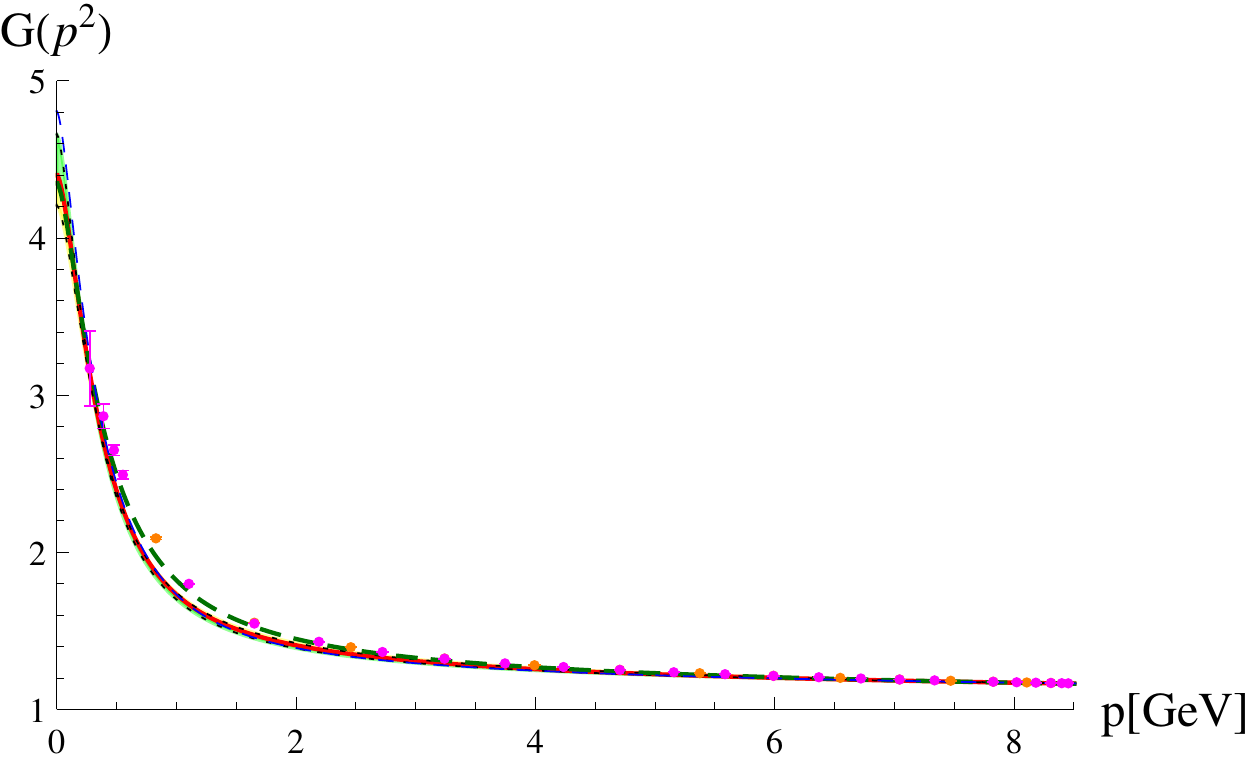}
  \caption{\label{fig:props}The $SU(3)$ gluon (left) and ghost (right) dressing functions from different truncations compared to lattice data \cite{Sternbeck:2006rd}. The input is depicted by a blue/dashed line, the results using the calculated three-gluon vertex $D^{A^3}_o$ by a red/continuous line and a band and the results from the three-point closed truncation by a green/dashed line.}
\end{figure}

\begin{figure}[tb]
  \includegraphics[width=0.45\textwidth]{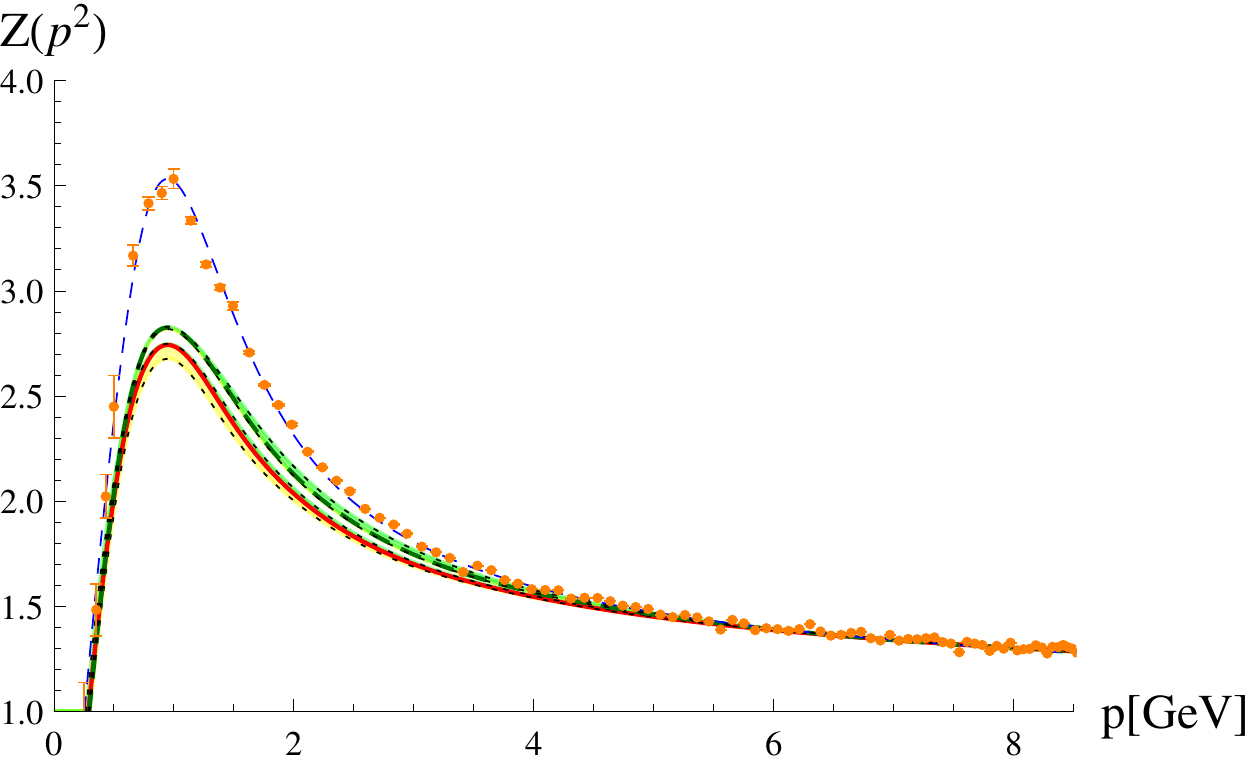}
  \hfill
\includegraphics[width=0.45\textwidth]{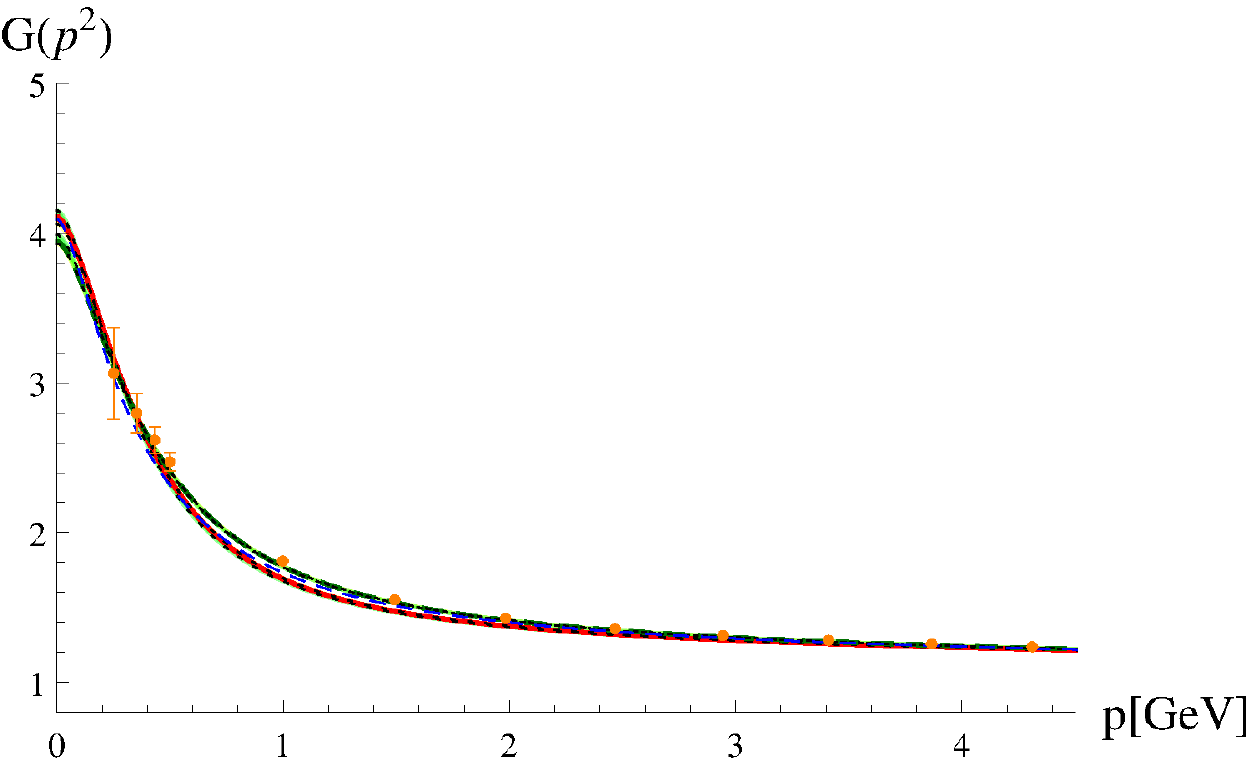}
  \caption{\label{fig:props-SU2}The $SU(2)$ gluon (left) and ghost (right) dressing functions from different truncations as in fig.~9. In addition the results from the three-point closed truncation (upper line) also have a thin band corresponding to different three-point vertices. Lattice data is from ref.~\cite{Sternbeck:2007ug}.}
\end{figure}

The projection used for the three-gluon vertex has the advantage that it corresponds exactly to the expression of the three-gluon vertex that appears in the DSE for the gluon propagator. From the comparison with lattice data we can infer that any errors we made by neglecting tensors and further diagrams is small, at least for this specific projection. Thus, although we used a truncated DSE, we can use the results for the three-gluon vertex from Sec.~\ref{sec:3g} in the gluon loop and assume that our errors are small compared to the only other remaining source of errors, viz. the neglected two-loop diagrams. The ghost-gluon vertex is as before taken from Sec.~\ref{sec:props+ghg}. Having a band of results for the three-gluon vertex we can even estimate how large the effect of any change in the vertex is for the gluon dressing, see  \fref{fig:couplings} and \fref{fig:props}. A further improvement of the lattice data would allow to narrow down the band. But even with the current width we conclude that the impact of two-loop diagrams is not negligible. We performed the same calculation for the $SU(2)$ propagators, see \fref{fig:props-SU2}. As ghost-gluon vertex we used in this case the result obtained from calculating the vertex with the propagators discussed in Sec.~\ref{sec:props+ghg}.

\section{Dynamical calculation of two- and three-point functions}
\label{sec:selfConsistent}

Finally we solve the full system of propagators and three-point functions. The only required input for this system is the model for the four-gluon vertex for which we use the same one as for the three-gluon vertex-only calculation with the parameters corresponding to the central line in the band of solutions. In figs.~\ref{fig:3g_singleScale} and \ref{fig:props} the green, dashed lines represent the solution from the fully coupled system. Clearly, the three-gluon vertex is too strong at low momenta. This is a direct consequence of the missing strength in the gluon dressing. Within our truncation this can only be improved by adding the two-loop terms in the gluon propagator DSE. When the gluon dressing moves up, the three-gluon vertex dressing must automatically come down in the IR.
Also in the plots for $SU(2)$, figs.~\ref{fig:3g_singleScale-SU2} and \ref{fig:props-SU2}, green, dashed lines show the solution from a self-consistent calculation. In addition we varied the parameters for the four-gluon vertex model as in the three-gluon vertex-only calculation. However, the effect is smaller and hardly visible at all in the propagator dressings.

\section{Summary and conclusions}

We presented results for the propagators of Landau gauge Yang-Mills theory. Using an effective model for the three-gluon vertex we could obtain good agreement with lattice results \cite{Huber:2012kd}. The obtained propagators formed the basis to calculate the three-gluon vertex. By varying the parameters of the four-gluon vertex, the only remaining model input, we produced a band of solutions that cover the available lattice data. We could use the three-gluon vertex results to show that the contribution from two-loop diagrams is not negligible \cite{Blum:2014gna}. As a consequence, any truncation scheme aiming at a quantitative improvement of the gluon propagator must include the two-loop diagrams. We tested explicitly that a self-consistent solution of two- and three-point functions \cite{Blum:2014gna} does not alleviate this problem and due to the missing strength in the gluon propagator also the results for the ghost-gluon and three-gluon vertices deviate from available lattice results. On the other hand, taking propagators that are in good agreement with lattice results the corresponding three-point functions can be reproduced rather well. Hence neglected diagrams seem to play a smaller role for vertices than for the gluon propagator which is encouraging evidence for the convergence of this kind of vertex expansion of QCD.

\section*{Acknowledgments}
M.Q.H. and L.v.S. thank the organizers for a stimulating and pleasant workshop QCD-TNT-III.
This work was supported by the Helmholtz International Center for FAIR within the LOEWE program of the State of Hesse, the European Commission, FP7-PEOPLE-2009-RG No. 249203, the Alexander von Humboldt foundation and the BMBF grant OSPL2VHCTG. Feynman diagrams were created with \textit{FeynDiagram} and \textit{JaxoDraw} \cite{Binosi:2003yf}.

\bibliographystyle{utphys_mod}
\bibliography{literature_QCD_TNTIII}

\end{document}